\documentstyle[a4p,12pt,cite,psfig]{article}
\begin{document}
 
\begin{titlepage}
\vspace{4mm}

\begin{flushright}
TAUP-2563-99 \hspace*{5mm} \\
WIS-99/04/Feb.DPP
\end{flushright}
\vspace{11mm}
\flushbottom
%
\vspace{8.mm}

\begin{center}
{\large\bf Isospin Invariance and Generalized Bose Statistics}
\end{center}
\begin{center}
{\large\bf applied to Low Energy {\boldmath $K^{\pm}K^o$} and 
{\Large \boldmath $\pi^{\pm}\pi^o$}
Space Symmetries}
\end{center}
\vspace{0.4cm}
 
\begin{center}
{\large G}{\small IDEON} {\large A}{\small LEXANDER}$^{a,}$\footnote{\tt
alex@lep1.tau.ac.il} 
\ and \
{\large H}{\small ARRY} {\large J}. {\large L}{\small IPKIN}$^{a,b,}$\footnote{\tt ftlipkin@wiswic.weizmann.ac.il}
\vspace{5mm}
 
a) School of Physics and Astronomy\\
Raymond and Beverly Sackler Faculty of Exact Sciences\\
{\it Tel-Aviv University}\\
Tel-Aviv 69978, Israel\\
\vspace{1.5mm}
 
b) Department of Particle Physics\\
{\it Weizmann Institute of Science}\\
Rehovot 76100, Israel\\
\vspace{6mm}
 
\today\\
\end{center}
\vspace{1.2cm}
 
\begin{abstract}
The use of isospin invariance and Generalized Bose statistics shows that the
Bose-Einstein correlation (BEC) of identical bosons seen in the $K^+K^+$ and 
$\pi^+ \pi^+$ systems can be extended to apply to $K^+ K^o$ 
and  $\pi^+ \pi^o$ pairs when they are produced from an initial state with 
isospin zero; e.g. by the fragmentation of gluons or of a strange quark pair
accompanied by gluons. 
This might be useful in constructing more realistic model dependent Monte
Carlo programs for the investigation 
of the isospin structure of the particles produced in a given 
kinematic region like the central region in high energy hadronic collisions
or quark-gluon plasma. 
Some of the consequences of this extension are
here applied to the hadronic $Z^o$ decays. In particular the study
addresses the question  
how much of the observed low  $K^o_S K^o_S$ mass enhancement is to be
attributed to a BEC effect and how much to the
$f_0(980)\to K^o_S K^o_S$ decay.
Finally we point out the restrictions imposed by the extension
on the choice of reference samples for
the BEC studies.

\end{abstract}
\vspace{5mm}
\begin{center}
(Submitted for publication)
\end{center}
\end{titlepage}
\newpage
\section{Introduction}
\label{intro}
 
\ \ \ \ \ In reactions between particles which lead to
multi-hadronic final states the constructive interference between two
identical bosons, the so called
Bose-Einstein Correlation (BEC), is well known.
These correlations lead to an enhancement of
the number of identical bosons over that of
non-identical bosons when
the two particles are close to each other in phase space.
Experimentally this effect, also known as the GGLP effect,
was for the first time observed in particle physics
by Goldhaber et al.
\cite{goldhaber} in like-sign charged pions produced in 
$\overline{p} p$
annihilations at $\sqrt{s}$ = 2.1 GeV.
In addition to the quantum mechanical aspect of the BEC,
these correlations are also
used to estimate the dimension of the emitting source
of the identical bosons \cite{bowlers}. Recently the interest in 
the BEC of identical charged pions has been extended 
to their possible effect on the mass measurements of the $W$ gauge bosons
in the the reaction $e^- e^+ \to W^+ W^-$ which has been
dealt with theoretically \cite{lundww} and experimentally \cite{expww}. 
Finally in high energy 
reactions the BEC of charged pions may influence the properties of the 
multi-hadron final states. Therefore more efforts are now devoted to
estimate these effects by incorporating BEC in the various 
model based Monte Carlo programs that are confronted with the data.\\
 
The relation to the size of the emission region, which is often 
assumed to be spherical with a Gaussian distribution, is given by the
well known  
formula for the  correlation function $C_2$ of two identical particles with four
momenta $q_i \ ( i = 1, 2 )$ and $Q\ = \ \sqrt{-(q_1 - q_2)^2}$ namely,
\begin{equation}
\label{BOWLER}   
\sigma_{tot} \frac{d^2 \sigma_{12}}{d \sigma_1
\,d\sigma_2}\,\,\equiv\,\,C_2(Q)\,\,\,=\,\,1\,\, +\,\,
\lambda e^{-r^2Q^2}\,\, ,
\end{equation}
where $\lambda$, the chaoticity parameter which can vary between
0 and +1, measures the strength
of the effect.
The term $e^{-r^2Q^2}$  is the
normalised Fourier transform of $\rho$, the source density.
Some recent reviews which summarize
the underlying BEC theoretical aspects and the
experimental results are given, for example, in references
\cite{hofmann,marcellini,wolf}.\\
 
It has further been shown, that under certain conditions,
a Bose-Einstein like enhancement can also be expected in a
boson-antiboson system like the $K^o \overline{K}^o$ pair
\cite{alex1,lipkin} if the $\lambda$ parameter is
larger than zero.
In fact, in
a sample of spinless boson-antiboson system, which is a mixture of
C = +1 and C = --1 states, the C = +1 part behaves like identical
bosons, that is, it produces a BEC like low mass enhancement
whereas the C = --1 part
decreases to zero as $Q_{K,\overline{K}} \to 0$. 
This has been experimentally demonstrated by
the OPAL collaboration \cite{opalk0k0} in their study of the
$K^o_S K^o_S$ system
and was later
confirmed by the DELPHI \cite{delphi} and ALEPH \cite{aleph}
groups at LEP. This GGLP low mass enhancement is the result of the
spatial symmetric state which is always the case for identical Bosons
and is also true for the  C = +1 part of the Boson-Antiboson
system.\\

Recently it has been pointed out \cite{alexlipkin} that
a pair of identical fermions, $ff$, having a total spin of $S$ = 1,
will have an $Q_{ff}$ dependence near threshold
similar to that of a Boson-Antiboson pair with the eigenvalue of
C = --1. This follows from the Pauli exclusion principle and
the expectation that
in the absence of low mass p-wave resonances
the relative contribution of the $\ell = 0$ state
increases as $Q_{ff}$ decreases due to the angular momentum barrier.
The measurement of this behavior allowed to estimate for
the first time the
dimension $r$ of the identical fermions $\Lambda \Lambda 
(\bar \Lambda \bar \Lambda)$ 
emitter \cite{lamtotal}
in a similar way as
it has been done in the past for identical bosons emitter.\\

In the same way that the Pauli principle can be used to treat particles in
the same spin multiplet to show BEC-like effects,
and is generalised to treat particles in the same isospin multiplet
one can also consider using the 
generalised Bose statistics to treat
hadrons in the same isospin multiplets as being identical.\\

Bose statistics allows only even partial waves for states of two identical
spinless bosons. 
Similarly generalized Bose statistics allows only even partial 
waves for states of two spinless bosons in the same isospin multiplet if they
are in an isospin eigenstate which is symmetric under 
isospin permutations; e.g.
I=1 for two kaons or I=0 or 2 for two pions. Thus pairs of bosons which are in
these symmetric isospin eigenstates can be expected to show the
analogue 
of Bose 
enhancement. These effects should be easily observable in cases where kaon or
pion pairs are produced inclusively from a initial isoscalar state. In this
case we shall see that the s-wave amplitudes for $K^+ K^o$ and $\pi^+ \pi^o$
states should be exactly the same as those for the corresponding identical
boson states $K^+ K^+$ and $\pi^+ \pi^+$ produced in the same
experiment\footnote{Whenever we refer to a specific two-boson state
we also mean its charge conjugate one.}.\\ 

Here we discuss the consequences of a generalised Bose
statistics and isospin invariance to the properties of two-boson
systems belonging to the same isospin multiplet which emerge from an
I=0 state. As an example
we address our study to the $K^+ K^o$ and $\pi^+ \pi^o$ pairs 
present in the hadronic $Z^o$ decays. In Section 2 we 
we present the
basic relations between pairs of kaons obtained from the generalised
Bose statistics. Section 3 deals specifically with question concerning
the interpretation of the observed $K^o_S K^o_S$ low mass enhancement
in multi-hadronic final states.
In Section 4 we deal with the two-pion system and 
in Section 5 we discuss the possible deviations from an isoscalar dominance.
Finally a summary and conclusions are presented in Section 6.

\section{Isospin Invariance applied to hadrons produced from an 
I=0 state}
\label{Isoscalar}

In many cases boson pairs, like the $KK$ and $\pi\pi$ systems, 
are produced from an initial state which is
isoscalar to a very good approximation. Among them are pairs produced from
an initial multi-gluon state as in the central region of a high energy
collision, pairs from hadronic decays of isoscalar heavy quarkonium resonances;
e.g. $J/\psi$ or $\Upsilon$, or pairs produced by $Z^o$ decays. In some of these
cases the initial state is however not pure I=0 but has also 
some contamination of an
I=1 component which is mixed in like in those processes  
where the $J/\psi$ and $\Upsilon$  
decay to hadrons via  one photon annihilation.
According to the specific case methods can be applied to reduce, or even
to eliminate, this contamination. For example, the subsample of the
C = --1 quarkonia which decay into an odd number of pions assure, 
due to G-parity,
that the hadronic final state is in an I=0 state. 
This method is not useful for the BEC
measurements of the hadronic $Z^o$ decays. 
Multi-hadron final states however which originate 
from the $Z^o$ decay to the heavy quarks, $s \bar s,\ c \bar c $ and 
$b \bar b$, are  
in an I=0 state. 
An efficient signature for a $Z^o \to s \bar s$ decay produced
in $e^+ e^-$ collisions is for example 
given by the hadronic final states which contain a high momentum
$\Lambda$ \cite{gustaf}. In fact for $Z^0$ decay events with a
$\Lambda$ momentum  $P_{\Lambda}/P_{beam} > 0.4$ the fraction of I=0
may amount to more than 70$\%$. \\ 

In the following we restrict our study to the
three di-kaon
systems with strangeness +2 namely, $K^+K^+$, $K^oK^o$
and $K^+K^o$ pairs, which are part of a   
multi-particle final state produced from an isoscalar state.  
We note that $K^+K^+$ and $K^oK^o$ are pure I=1 isospin states, while the
$K^+K^o$ is a mixture of two isospin states, I=1 and I=0. Bose statistics 
further tells us that in the KK center-of-mass 
system the $K^+K^+$ and $K^oK^o$ are built of
two 
identical particles and therefore have only even partial waves. The $K^+K^o$ 
has both even and odd partial waves, because these two kaons are not 
identical. The generalised Bose statistics however tells us that the I=1 state
of this system has only even partial waves while the I=0 state has only odd
partial waves.\\ 

Since the $K^+K^+$ state has
isospin quantum numbers $(I,I_z) = (1,+1)$ the accompanying multi-particle
state 
is required by isospin invariance to be an 
isospin eigenstate with the isospin quantum numbers $(I,I_z) = (1,-1)$
which we denote as $X_{1,-1}$.
Isospin invariance further requires that the other multi-particle states in the
same isospin multiplet denoted by $X_{1,-I_z}$ be equally produced together
with the two-kaon multiplet carrying the quantum numbers $(1,I_z)$. 
We can therefore write the following relation between the amplitudes for
production of these states from an initial isoscalar state, denoted by $i_o$,
$$\sqrt {2} \cdot A[i_o \rightarrow K^+(\vec p) K^+(-\vec p) X_{1,-1}] = 
\sqrt {2} \cdot A[i_o \rightarrow K^o(\vec p) K^o(-\vec p) X_{1,+1}] = $$
$$ = -A[i_o \rightarrow K^+(\vec p) K^o(-\vec p) X_{1,0}] -  
A[i_o \rightarrow K^o(\vec p) K^+(-\vec p) X_{1,0}]\ ,     \eqno(2)           
$$
where $\vec p$ denotes the momentum of the kaon in the centre 
of mass system of
the two kaons.\\
 
There is an additional isospin-zero amplitude for the 
$K^+K^o$ final state which
can be produced together with an isoscalar multi-particle state 
which we denote by $X_{0,0}$. 
Since the generalized Bose statistics requires the
isoscalar state to be antisymmetric in space, the amplitude for the production
of this state satisfies the relation:
$$A[i_o \rightarrow K^+(\vec p) K^o(-\vec p) X_{0,0}] = - 
A[i_o \rightarrow K^o(\vec p) K^+(-\vec p) X_{0,0}]\ .     \eqno(3)           $$
This isoscalar contribution clearly vanishes in the limit 
$\vec p \rightarrow 0$ which is relevant for the BEC. Since the multi-particle 
states $X_{1,0}$ and $X_{0,0}$ accompanying the kaon pair have different 
isospin, they are orthogonal and all interference terms between the two 
isospin amplitudes must cancel if there is no measurement on the accompanying 
multi-particle state. Plots of the number of pairs versus Q should be identical
for $K^oK^+$, $K^oK^o$  and $K^+K^+$ in the low Q region where only s-waves
contribute. This means that if a BEC enhancement is seen in the
$K^+K^+$ it should also be present in the $K^oK^+$ and the $K^oK^o$ systems. 
At higher energy, as soon as p-waves can contribute to 
$K^oK^+$ but not of course to
$K^+K^+$ nor $K^oK^o$, there should be an excess of events for $K^oK^+$.\\ 

We now note that the $K^o$ states are in general not detected as such but
rather as $K^o_S$ and $K^o_L$, which are mixtures of $K^o$ and $\bar K^o$. The
contribution from the $K^o$ component is well defined and satisfies the
isospin predictions. However, the contribution from the $\bar K^o$ component is
completely unrelated except for the fact that it is positive definite and thus
the isospin relations provide lower bounds. A detailed analysis is given below
for the $K^o_S K^o_S$ system where the simplifications from the Bose symmetry
of the identical particles allow the testing of reasonable assumptions for the
additional $K^o\bar K^o$. contribution. The $K^+ K^o_S$ system is more
complicated because of the additional unknown contributions from odd partial
waves, and is not treated further here. 

\section{The $K^o_S K^o_S$ system}
 
The $K^o_SK^o_S$ system has only even partial waves and is a linear combination
of $K^oK^o$ and $K^o\bar K^o$. The $K^oK^o$ component is related to the
$K^+K^+$ component by the isospin relation (2). We can thus write the following
relations for the probabilities, denoted by P, for the detection of these
states. 

$$ P[i_o \rightarrow K^o(\vec p) K^o(-\vec p) X_{1,+1}
\rightarrow K^o_S(\vec p) K^o_S(-\vec p) X_{1,+1} ] = $$
$$ (1/2)\cdot P[i_o \rightarrow K^o(\vec p) K^o(-\vec p) X_{1,+1}] = 
(1/2)\cdot P[i_o \rightarrow K^+(\vec p) K^+(-\vec p) X_{1,-1}]\ , 
\eqno(4)           
$$
where the factor(1/2) arises from the division of $K^oK^o$
into $K^o_LK^o_L$ and $K^o_SK^o_S$.\\ 

In a realistic experiment the final $K^o_S K^o_S$ is identified but the
specific multi-particle state $X_{I,I_z}$ is not and the result is 
obtained by summing over all
possible multi-particle final states here denoted simply by $X$. 
Since these states include  
both states of strangeness 
$\pm 2$ and strangeness 0, the final $K^o_S K^o_S$ pairs
included in the sum may come not only from $K^o K^o$ and 
$\bar K^o \bar K^o$ states but also the even parity $\bar K^o K^o$ and
$K^o \bar K^o $. Thus we can write 
$$ \sum_X P[i_o \rightarrow K^o_S(\vec p) K^o_S(-\vec p) X] = $$
$$(1/4) \cdot \sum_X P[i_o \rightarrow K^+(\vec p) K^+(-\vec p) X]
+  (1/4) \cdot \sum_X P[i_o \rightarrow K^-(\vec p) K^-(-\vec p) X]  $$
$$ + \sum_X P[i_o \rightarrow K^o \bar K^o X
\rightarrow K^o_S(\vec p) K^o_S(-\vec p)X]\ , 
\eqno(5a)           
$$
where $P[i_o \rightarrow K^o \bar K^o X \rightarrow K^o_S(\vec p) 
K^o_S(-\vec p)X]$ 
denotes the sum of the probabilities of all transitions from the initial state 
$i_o$ to the final state $K^o_S(\vec p) K^o_S(-\vec p)X$ 
via any $K^o \bar K^o X$
state. This can be conveniently rewritten
$$ {{\sum_X P[i_o \rightarrow K^o_S(\vec p) K^o_S(-\vec p) X]}\over{
\sum_X P[i_o \rightarrow K^+(\vec p) K^+(-\vec p) X]
+ \sum_X P[i_o \rightarrow K^-(\vec p) K^-(-\vec p) X]}}  
= $$

$$= {{1}\over{4}}
+  {{\sum_X P[i_o \rightarrow K^o \bar K^o X
\rightarrow K^o_S(\vec p) K^o_S(-\vec p)X] }\over{
\sum_X P[i_o \rightarrow K^+(\vec p) K^+(-\vec p) X]
+ \sum_X P[i_o \rightarrow K^-(\vec p) K^-(-\vec p) X]}}\ . 
\eqno(5b)           $$

This last relation has a bearing on the analysis of the enigmatic 
$f_o(980)$ scalar resonance which has a long history regarding its nature
and decay modes \cite{morgan}. Its existence is well established
by its decay into a pair of pions. As for its decay 
into the $K\bar K$ final states the situation is more complicated.
The $f_o(980)$
central mass value   
lies below the $K\bar K$ threshold however the upper part of its width is 
above it. At the same time the $f_o(980) \to K\bar K$ decay branching is a major
tool in nailing down the nature of this  resonance and its total
width. Since the analyses of the final 
$K^+ K^-$ state are handicapped by the strong presence of the $\phi(1020)
\to K^+ K^-$ decay many of the analyses utilised instead
the $K^o_S K^o_S$ final state system. The origin of the excess of
$K^o_S K^o_S$ near threshold however could a priori have two sources,
the decay product of the $f_o(980)$ resonance and the 
BEC enhancement.\\ 

Since the denominators of Eq. (5b) come from exotic kaon pair states which
have no resonances, the presence of resonances in the numerator of the second 
term on the right hand side of (5b) will show up as energy-dependent enhancements over the 
background which is expected to be similar to that of the first term, namely 
$\approx 1/4$. Note that the observed number of counts for decays into the   
$K^o_S K^o_S$ mode is reduced by an additional factor of (4/9) because 
in general only the 
$\pi^+\pi^-$ decay mode of each $K^o_S$ is detected. Thus the statistical errors
in the denominators of Eq. (5b) are expected to be much lower than those of the
numerators.\\ 

Eqs. (5) relate the BEC excess of the  $K^o_S K^o_S$
to that of the
$K^+ K^-$ system which, for example, 
was measured and found to exist    
in the hadronic $Z^o$ decay events \cite{delphikpkp}.
The experimental results of the low mass enhancement seen in the 
the $K^+ K^-$ and $K^o_S K^o_S$ systems determine 
the contribution from the $K^o \bar K^o$
states which may be taken as the  upper limit of
the $f_o(980) \to K^o \bar K^o$ decay rate.

\section{The generalised Bose statistics applied to pion pairs} 
As in the case of the kaon pairs, the production of $\pi^+\pi^+$ and
$\pi^+\pi^o$ pairs in a multi-particle final state can be related if
produced from an initial isoscalar state. 
Here $\pi^+\pi^+$ is a pure isospin state with I=2, while
$\pi^+\pi^o$ is a mixture of two isospin states with I=2 and I=1. Bose
statistics tells us that in the $\pi \pi$ centre-of-mass system the
$\pi^+\pi^+$ system has two identical particles and therefore has only even
partial waves. The $\pi^+\pi^o$ has both even and odd partial waves, because
these two pions are not identical. But the generalized Bose statistics tells us
that the I=2 state of this system has only even partial waves and the I=1 state
has only odd partial waves.\\ 
 
Since the $\pi^+\pi^+$ state has isospin quantum numbers $(I,I_z) = (2,+2)$ the
remaining multi-particle state which we denote by $X_{2,-2}$ is required by
isospin invariance to be an isospin eigenstate with the isospin quantum numbers
$(I,I_z) = (2,-2)$. Isospin invariance further requires that the other
multi-particle states in the same isospin multiplet denoted by $X_{2,-I_z}$ be
equally produced together with the two-pion multiplet carrying the quantum
numbers $(2,I_z)$ We can therefore write the following relation between the
amplitudes for production of these states from an initial isoscalar state,
denoted by $i_o$ 

$$\sqrt {2} \cdot A[i_o \rightarrow \pi^+(\vec p) \pi^+(-\vec p) X_{2,-2}] = 
-A[i_o \rightarrow \pi^+(\vec p) \pi^o(-\vec p) X_{2,-1}] -$$
$$ -A[i_o \rightarrow \pi^o(\vec p) \pi^+(-\vec p) X_{2,-1}]\ ,     \eqno(6) 
$$
where $\vec p$ is the momentum of the pion in the centre of mass system of
the two pions.\\
 
There is an additional isospin-one amplitude for the $\pi^+\pi^o$ final state 
which can be produced together with an I=1 multi-particle state which we denote 
by $X_{1,-1}$. Since the generalized Bose statistics requires the I=1
state to be antisymmetric in space, the amplitude for the production
of this state satisfies the relation:
$$A[i_o \rightarrow \pi^+(\vec p) \pi^o(-\vec p) X_{1,-1}] = - 
A[i_o \rightarrow \pi^o(\vec p) \pi^+(-\vec p) X_{1,-1}]\ .     \eqno(7)           $$
This I=1 contribution clearly vanishes in the limit 
$\vec p \rightarrow 0$ which is relevant for BEC. The I=2 and I=1 amplitudes 
have opposite parity. Again the interference terms between the two isospin 
amplitudes cancel out if there is no measurement on the accompanying 
multi-particle state.
Plots of the number of events versus Q should be identical for $\pi^o\pi^+$ and 
$\pi^+\pi^+$ in the low Q region where only s-waves contribute. As soon as 
p-waves can contribute to $\pi^o\pi^+$ but not of course to $\pi^+\pi^+$ there 
should be an excess of events for $\pi^o\pi^+$. One can expect a large p-wave
contribution because of the presence of the $\rho$ resonance, and the tail of
the $\rho$  may still be appreciable at the $\pi \pi$ threshold.\\

In the $\pi \pi$ system there are more states and more isospin amplitudes than
in the KK system. There are also the $\pi^+\pi^-$ and $\pi^o \pi^o$ states and
an additional I=0 amplitude. The $\pi^o \pi^o$ state has two identical particles
and only even partial waves in the $\pi \pi$ center of mass system. But it is a
linear combination of two isospin states, I=0 and I=2, and therefore is not 
related directly to the even partial waves of the I=2 system. The $\pi^+\pi^-$ 
state has all three isospin eigenvalues, 0, 1 and 2, and both even and odd 
partial waves. The odd partial wave amplitudes have I=1 and are directly related
to the odd partial waves of the $\pi^o\pi^+$ system. The even partial waves 
have both I=0 and I=2 components and can be related to the other I=2 and I=0 
amplitudes by a a full amplitude analysis. In the low Q region where only 
s-waves contribute the $\pi^+\pi^+$, $\pi^+\pi^-$ and $\pi^o \pi^o$ amplitude 
depend upon only two isospin amplitudes, I=0 and I=2. Their intensities in this
region satisfy a triangular inequality
$$  
\sum_X\left |~\sqrt{
(2/3)\cdot P[i_o \rightarrow (\pi^o\pi^o) X]} - 
\sqrt{ (1/3)\cdot P[i_o \rightarrow (\pi^+\pi^-)_e X]}~\right | \leq  
$$ $$ \leq                            
\sum_X\sqrt{ P[i_o \rightarrow (\pi^\pm\pi^\pm) X]} =
\sum_X\sqrt{ P[i_o \rightarrow (\pi^\pm\pi^o)_e X]} \leq
$$ $$ \leq                            
\sum_X\left |~\sqrt{(2/3)\cdot P[i_o \rightarrow (\pi^o\pi^o) X]} + 
\sqrt{ (1/3)\cdot P[i_o \rightarrow (\pi^+\pi^-)_e X]}~\right |\ , 
\eqno(8) $$
where the notation $(\pi\pi)_e$ is used to indicate that only even
partial waves for the $(\pi\pi)$ final state are included in the sum.

\section{Possible deviations from isoscalar dominance} 

Precise calculations of the errors in the predictions based on an assumed
charge-symmetric or isoscalar final state are difficult. We present here some
arguments supporting the neglect of deviations from symmetry. But rather than
trying to improve these arguments it seems more profitable to test these
predictions by experiment. Since the isospin predictions are energy
independent, they can be tested over an energy range which includes not only
the Bose-Einstein and resonance regions but also higher relative momenta where
both these effects should be absent.\\ 

In processes leading to multi-particle states, nearly all
of the final quarks and antiquarks are produced by isoscalar gluons and this 
should enhance isoscalar dominance. For example, the charge asymmetry in a 
fragmentation process initiated by a $u \bar u$ pair produced in a $Z^o$ decay 
can easily be washed out quickly in a fragmentation process dominated by 
isoscalar gluons. We present two examples of how this can occur.\\ 

When a leading u-quark in the final stage of the fragmentation process picks
up a strange antiquark, the spins of the quark and antiquark are expected to be
uncorrelated. This gives a 3:1 ratio favoring $K^{*+}$ production over 
$K^+$. The final states $K^+$, $K^+\pi^o$ and $K^o\pi^+$ 
are then produced with 
the ratio 1:1:2 giving equal probabilities of producing a $K^+$ and a $K^o$ in 
the final state. The charge asymmetry is completely lost in this approximation
where effects of the $K-K^*$ mass difference are neglected.\\

If the leading u-quark in the final stage of the fragmentation process picks up
a nonstrange antiquark from a $(u \bar u) $ or $ (d \bar d) $ pair created with
equal probability by gluons, the leading u-quark will combine with the $ \bar u
$ or $ \bar d $ to make a final nonstrange meson, leaving the remaining $u$ or
$d$ to continue the fragmentation process with equal probabilities. 
   $$ u + (u \bar u) \rightarrow M^o(u \bar u) + u           \eqno (9a)  $$
   $$ u + (d \bar d) \rightarrow M^+(u \bar d) + d           \eqno (9b)  $$
where $M^o$ and $M^+$ denote neutral and positive final meson states.  Thus the
initial charge asymmetry leaves the process as a charge asymmetry between the
$M^o$ and $M^+$ mesons and the remaining quark which continues the 
fragmentation process is charge symmetric.\\ 

These two mechanisms for washing out the charge asymmetry by gluons assume 
mass degeneracies whose breaking can introduce charge asymmetry. The 
$K-K^*$ mass difference suppresses $K^*$ production and the $d$ quark 
production in the final state. But the $\eta-\eta'-\pi$ mass differences 
suppress the reaction (9a) while leaving (9b) unaffected and thus suppress
$u$ quark production. Thus these two symmetry-breaking mechanisms work in 
opposite directions and the violations of charge symmetry
observed in the total final state are expected to be small.\\ 

We also note that in $Z^o$ decay one can expect that the isoscalar final states
arising from initially produced $s \bar s$, $c \bar c$ and $b \bar b$ pairs will
occur at least equally with the production of $u \bar u$ and $d \bar d$ pairs,
This gives an additional factor of two favoring isoscalar states over those 
discussed above.  
 
\section{Summary and conclusions}
From the generalised Bose statistics and isospin invariance follows
that also boson pairs with different charges which belong to the same
isospin multiplet may show low mass BEC enhancement. In the case that
these pairs, together with their accompanying hadrons, are produced from
a pure I=0 state, relations can be derived between their low energy
production amplitudes
and those of two identical bosons in the same isospin multiplet. In
verifying this effect experimentally, e.g. in the hadronic $Z^o$
decays, one should however remember that the condition for a pure I=0 state
is not 100$\%$ satisfied. Furthermore, pions originating from weak or 
electro-magnetic decays, such as the $\eta$ decay or from unresolved
$K^o_S$ decays, may also affect slightly the relation given in Eq. (6).\\

In as much that BEC effects of identical boson are relevant
for the description of the data in terms of model dependent 
Monte Carlo program,
one should also consider these effects in the 
$\pi^{\pm} \pi^o$ and $K^+ K^o (K^- \bar K^o)$ systems. 
In the experimental studies of the BEC the 
choice of the reference sample against which the effect is
measured is crucial. Ideal reference samples are those which retain all
the features and correlations of the data 
apart from the one due to the BEC. 
The relations given by the generalised Bose statistics
restrict the choice of these reference samples. For example,
for the study of the BEC of the $\pi^{\pm} \pi^{\pm}$ pairs
a data reference sample constructed out of  $\pi^{\pm} \pi^o$ 
is forbidden. Due to the triangle inequality given by Eq. (8),
it may be wise even to avoid the use of $\pi^+ \pi^-$ data pairs as a 
reference sample and to use instead a Monte Carlo generated sample.
Finally it is worthwhile to point out that in trying to 
associate the observed $K^o_S K^o_S$ low mass enhancement, seen in
multi-hadron final states, to the
$f_o(980) \to K^o_S K^o_S$ decay channel one should also 
examine the possible contribution from the BEC to this final state.
  

\end{document}